\documentclass[12pt]{article}
\usepackage[latin1]{inputenc}
\usepackage[british]{babel}
\usepackage{cmap}
\usepackage{lmodern}

\usepackage{amssymb, amsmath, amsthm}
\usepackage[a4paper,top=25mm,bottom=25mm,left=25mm,right=25mm]{geometry}
\usepackage{ragged2e}

\usepackage{authblk} 
\usepackage{pifont}
\usepackage{graphicx}
\usepackage[usenames,dvipsnames,svgnames,table]{xcolor}
\usepackage[figuresright]{rotating}
\usepackage{xtab} 
\usepackage{longtable} 
\usepackage{multirow}
\usepackage{footnote}
\usepackage[stable]{footmisc}
\usepackage{chngpage} 
\usepackage{pdflscape} 
\usepackage[nottoc,notlot,notlof]{tocbibind} 

\usepackage{pgfplots}
\pgfplotsset{compat=1.14}
\pgfplotsset{every tick label/.append style={font=\footnotesize}}
\usepgfplotslibrary{fillbetween}
\usepackage{setspace}

\makesavenoteenv{tabular}
\usepackage{tabularx}
\usepackage{booktabs}
\usepackage{threeparttable} 
\usepackage[referable]{threeparttablex} 
\newcolumntype{R}{>{\raggedleft\arraybackslash}X}
\newcolumntype{L}{>{\raggedright\arraybackslash}X}
\newcolumntype{C}{>{\centering\arraybackslash}X}
\newcolumntype{M}[1]{>{\centering\arraybackslash}m{#1}}
\newcolumntype{K}{>{\columncolor{gray!20}}C}
\newcolumntype{k}{>{\columncolor{gray!20}}c}

\usepackage{dcolumn} 
\newcolumntype{.}{D{.}{.}{-1}}

\usepackage{tikz}
\usetikzlibrary{arrows, calc, matrix, patterns, positioning, trees}
\usepackage[semicolon]{natbib}
\usepackage[hyphens]{url}
\usepackage{hyperref} 
\hypersetup{
  colorlinks   = true,    
  urlcolor     = blue,    
  linkcolor    = blue,    
  citecolor    = ForestGreen      
}
\usepackage{microtype}
\usepackage[justification=centerfirst]{caption} 

\usepackage[labelformat=simple]{subcaption}

\DeclareCaptionLabelFormat{parenthesis}{(#2)}
\makeatletter
\renewcommand\p@subfigure{\arabic{figure}.}
\makeatother

\DeclareCaptionLabelFormat{parenthesis}{(#2)}
\makeatletter
\renewcommand\p@subtable{\arabic{table}.}
\makeatother

\usepackage{enumitem}
\setlist[itemize]{leftmargin=2.5\parindent}
\setlist[enumerate]{leftmargin=2.5\parindent}

\def\addlegendimage{\csname pgfplots@addlegendimage\endcsname}

\theoremstyle{plain}

\theoremstyle{definition}

\theoremstyle{remark}

\def\keywords{\vspace{.5em} 
{\noindent \textit{Keywords}: }}

\def\JEL{\vspace{.5em} 
{\noindent \textbf{\emph{JEL} classification number}: }}

\def\AMS{\vspace{.5em} 
{\noindent \textbf{\emph{MSC} class}: }}

\title{The unfairness of the revenue distribution \\ system used in the UEFA club competitions}
\author{\href{https://sites.google.com/view/laszlocsato}{L\'aszl\'o Csat\'o}\thanks{~E-mail: \emph{laszlo.csato@sztaki.hu}} }
\affil{Institute for Computer Science and Control (SZTAKI) \\
E\"otv\"os Lor\'and Research Network (ELKH) \\
Laboratory on Engineering and Management Intelligence \\
Research Group of Operations Research and Decision Systems}
\affil{Corvinus University of Budapest (BCE) \\
Department of Operations Research and Actuarial Sciences}
\affil{Budapest, Hungary}

\date{\today}

\def\Dedication{
{\noindent
``\emph{Given the objective function of the organizer and the technology of the auction/contest it is possible to design an optimal prize scheme contingent on the distribution of contestant abilities/willingness to pay.}''\footnote{~Source: \citet[p.~1137--1138]{Szymanski2003}.}
}

\flushright
\begin{small}
(Stefan Szymanski: \emph{The economic design of sporting contests})
\end{small}

\vspace{0.5cm} 
\justify }

\begin{document}

\maketitle
\thispagestyle{empty}
\Dedication

\begin{abstract}
\noindent
The Union of European Football Associations (UEFA) distributes commercial revenue from its club competitions according to a complex mechanism. Coefficient-based amounts are paid such that the participating teams are ranked on the basis of performances over the last 10 years. Due to the non-anonymity of this allocation rule, all clubs are interested in qualification together with lower-ranked teams, which makes the procedure incentive incompatible. In particular, the English club Arsenal is found to lose 132 thousand Euros prize money in the 2022/23 UEFA Europa League because of exerting full effort in its domestic championship. Two strategy-proof alternatives are proposed to divide fairly this pillar of the revenue distribution system.


\AMS{90B90, 91A80, 91B14}

\JEL{C44, D71, Z20}
\end{abstract}

\clearpage

\section{Introduction} \label{Sec1}

There are several views on what fairness in sports means \citep{Csato2021a, Pawlenka2005, Wright2014}. However, it is hard to debate that the prize allocation scheme should reward performance \citep{DietzenbacherKondratev2022}. Otherwise, a contestant might be strictly better off by losing, which can inspire tanking, the act of deliberately losing a game in order to gain other advantages.

Such behaviour threatens the integrity of sports, thus, academic research can make an important contribution by highlighting every issue of incentive incompatibility and making proposals to reduce or eliminate perverse incentives. International sports federations are strongly encouraged to take these findings into account as, according to the statement of the International Olympic Committee (IOC): international non-governmental organisations recognised by the IOC---among them the F\'ed\'eration Internationale de Football Association (FIFA)---``\emph{are responsible for the integrity of their sport on the international level}'' \citep{IOC2022}. Furthermore, since sport plays an important role in creating the cultures and identities of our communities \citep{HoulihanMalcolm2016}, the lack of integrity in sport might lead to negative social outcomes \citep{HoulihanDownwardYamamotoRasciuteTakasu2020, ManoliBanduraDownward2020}.

The business model of the sports industry relies on the players exerting their best efforts to win. Therefore, incentive incompatibility may seem to be merely a theoretical curiosity without any relevance in practice. However, many misaligned rules exist around the world. \citet{KendallLenten2017} offer probably the first comprehensive survey of them. \citet{LentenKendall2021} overview the problem of reverse-order player drafts. \citet{Fornwagner2019} presents field evidence that teams exploit the weakness of this draft system with a concrete losing strategy. The standard format of multi-stage contests with carried over results is also vulnerable to manipulation, and some historical examples are known when a handball team has not been interested in winning by a high margin \citep{Csato2022d}.

The Union of European Football Association (UEFA) has often used incentive incompatible rules. A team could have been strictly better off by losing in the qualification of the UEFA Europa League until the 2015/16 season \citep{DagaevSonin2018}, in the qualification of the UEFA Champions League between 2015/16 and 2018/19 \citep{Csato2019c}, and in the European qualifiers for the 1996 \citep{Csato2018c} and 2018 FIFA World Cups \citep{Csato2020c}. There has been a football match where winning was dominated by playing a draw for one team \citep[Section~2.1]{Csato2021a}, and another with both teams being interested in playing 2-2 \citep{Csato2020d}. The current seeding regime of the UEFA Champions League is not strategy-proof, namely, a team might be placed in a higher-ranked pot due to worse performance in its domestic league \citep{Csato2020a}. Finally, \citet{Csato2022a} demonstrates how the incentive incompatibility of the qualification systems for the 2020 UEFA European Championship \citep{HaugenKrumer2021} and the 2022 FIFA World Cup can be substantially improved by adding specific draw constraints to the set of restricted team clashes \citep{Kobierecki2022}.
These examples might suggest that European football has fundamental problems in contest design but they emerge mainly due to complex tournament structures and detailed regulations, which increase transparency compared to other sports.

This note will reveal a similar issue of incentive incompatibility in the current revenue distribution system of the UEFA club competitions.
The main novelties of the research can be summarised as follows:
\begin{itemize}
\item
Even though the effects of the revenue distribution system of the UEFA Champions League have already been investigated \citep{Bullough2018}, the theoretical properties of the allocation rule are first analysed here;
\item
The incentive incompatibility of the coefficient-based pillar is verified;
\item
Straightforward solutions are provided to eliminate tanking opportunities;
\item
In contrast to previous works, now the unique setting allows us to exactly quantify the financial consequences of misaligned sporting rules.
\end{itemize}
In particular, Section~\ref{Sec2} uncovers why the English club Arsenal has lost about 132 thousand Euros because it won against West Ham United on 1 May 2022. Section~\ref{Sec3} explores the root of the problem and suggests two alternatives to reform the revenue distribution system used by UEFA.

\section{A real-world example of unfair revenue allocation} \label{Sec2}

The commercial revenue from UEFA club competitions (UEFA Champions League, UEFA Europa League, UEFA Europa Conference League) is distributed to the clubs according to a complex scheme \citep{UEFA2022c}. First, some money is allocated to the teams eliminated in the qualifying phases as solidarity payment. The net amount available to the clubs that play in the group stage is divided into four pillars:
\begin{itemize}
\item
Starting fees: a guaranteed payment shared equally among the 32 participants;
\item
Performance-related fixed amounts: bonuses provided for wins and draws in matches played in the group stage, as well as for qualifying to a given stage of the tournament;
\item
Coefficient-based amounts: paid on the basis of performances over the last 10 years, including extra points for winning UEFA competitions in the past;
\item
Market pool: distributed in accordance with the proportional value of TV markets represented by the clubs.
\end{itemize}
All details can be found in a transparent format at \url{https://www.football-coefficient.eu/money/}.

We focus on the third pillar, the coefficient-based amounts, which sets out 30/15/10\% in the revenue distribution of the Champions League/Europa League/Europa Conference League, respectively. The 32 teams entering the group stage are ranked on the basis of the 10-year UEFA club coefficients. The lowest-ranked team receives one share and one share is added to every rank, so the highest-ranked club receives 32 shares \citep{UEFA2022c}.

The 10-year club coefficients used for the 2022/23 season of UEFA competitions are listed in \citet{Kassies2022b}. Among the 14 highest-ranked clubs, only Manchester United and Arsenal (both from England) have failed to qualify for the 2022/23 Champions League.

\begin{table}[t!]
\begin{threeparttable}
\centering
\caption{Final ranking of the top teams in the 2021/22 Premier League}
\label{Table1}
\rowcolors{1}{}{gray!20}
    \begin{tabularx}{\linewidth}{Cl CCC CCC >{\bfseries}C} \toprule \hiderowcolors
    Pos   & Team  & W     & D     & L     & GF    & GA    & GD    & Pts \\ \bottomrule \showrowcolors
    1     & Manchester City & 29    & 6     & 3     & 99    & 26    & $+73$    & 93 \\
    2     & Liverpool & 28    & 8     & 2     & 94    & 26    & $+68$    & 92 \\
    3     & Chelsea & 21    & 11     & 6     & 76    & 33    & $+43$    & 74 \\
    4     & Tottenham Hotspur & 22    & 5    & 11     & 69    & 40    & $+29$    & 71 \\
    5     & Arsenal & 22    & 3     & 13     & 61    & 48    & $+13$    & 69 \\
    6     & Manchester United & 16    & 10     & 12     & 57    & 57    & 0    & 58 \\
    7     & West Ham United & 16    & 8     & 14     & 60    & 51    & $+9$    & 56 \\ \toprule   
    \end{tabularx}
    
    \begin{tablenotes}
\item
\footnotesize{Pos = Position; W = Won; D = Drawn; L = Lost; GF = Goals for; GA = Goals against; GD = Goal difference; Pts = Points. All teams have played $38$ matches.}   
    \end{tablenotes}
\end{threeparttable}
\end{table}

Table~\ref{Table1} shows the final ranking of the 2021/22 English Premier League, which determines the qualification for the 2022/23 UEFA club competitions. Since Liverpool has won both the FA Cup and the EFL Cup that give a ticket to European competitions, the first four teams advance to the Champions League group stage, the fifth- and sixth-placed teams advance to the Europa League group stage, and the seventh-placed team goes to the Europa Conference League play-off round. Consequently, Arsenal and Manchester United are the highest-ranked clubs in the Europa League based on 10-year coefficients: Arsenal receives 31 and Manchester United receives 32 shares from the third pillar of the revenue distribution system.

Consider the counterfactual that Arsenal would have lost away against West Ham United on 1 May 2022 instead of winning. Then Arsenal would have had 66 and West Ham United would have had 59 points. Therefore, Arsenal and West Ham United would have qualified for Europa League, and Arsenal would have received 32 shares in the third pillar as being the highest-ranked club in this competition. According to the amounts distributed in the 2022/23 season (prior to COVID-19 impact deduction), one share corresponds to 132 thousand Euros here \citep{UEFA2022c}.

It can be checked that the amounts received from the other three pillars of the revenue distribution system (starting fees, performance-related fixed amounts, market pool) do not change if the rank of a team in its domestic league remains the same. Even though the split of the market pool between Arsenal and the other English club entering the Europa League is not specified exactly, this is clearly independent of the final ranking in the 2021/22 Premier League, which is not affected in our counterfactual.\footnote{~The regulation \citep{UEFA2022c} says that the cup winner receives 60\% and the other team 40\% but the split is equal (50\%-50\%) if the domestic cup winner does not qualify for the Europa League group stage. Therefore, it is not clarified whether Arsenal is counted as the cup winner instead of Liverpool or not.}
Consequently, Arsenal would have lost no money from the UEFA by losing against West Ham United.

The question arises: then why did Arsenal defeat West Ham United on 1 May 2022? At the moment of this game, Arsenal was still in contention for qualification for the Champions League, the more prestigious and lucrative European competition. However, one can easily devise a schedule where a misaligned incentive causes a real problem (see the case of the 2011/12 Dutch Eredivisie in \citet[Section~2.1]{Csato2021a}) and the unfair punishment of Arsenal can barely be debated.

\section{Discussion} \label{Sec3}

Owing to the misaligned design of the revenue distribution system used in UEFA club competitions, Arsenal has lost approximately 132 thousand Euros because it has won against West Ham United in the 2021/22 Premier League. This means an unfair punishment for better performance.

What is the problem behind the allocation of coefficient-based amounts?
The ranking of participating clubs in the three UEFA competitions does not satisfy \emph{anonymity}, that is, the share of any club depends on the identity of other qualified teams. Consequently, if the place of a team is already secured (see the robust margin of Arsenal over Manchester United in Table~\ref{Table1}), it is interested to qualify together with a lower-ranked club (West Ham United is preferred to Manchester United). That can be achieved in the domestic championship if at least two teams qualify, which holds---according to the 2022/23 UEFA access list \citep{UEFA2022b}---for all UEFA member associations except for Liechtenstein. Similar instances of financial losses are not ubiquitous only because the 10-year UEFA club coefficients are good predictors of the final ranking in the European football leagues.

The problem can be solved by an anonymous allocation rule that applies to all European competitions. Two alternatives are suggested for this purpose:
\begin{itemize}
\item
\emph{Rule A}: The coefficients-based amount is distributed on the basis of the position of the club among all clubs from its national association. \\
Formally, suppose that $k$ teams can qualify for \emph{all} European competitions from a given country. The highest $k$ coefficients of all teams in this association (including teams that do not qualify in the current season) determine the non-increasing vector $\mathbf{c}$ of financial coefficients $c_1 \geq \ldots \geq c_k$ of the country. The team ranked $\ell$th among the $k$ qualified teams receives the prize associated with the financial coefficient $c_{\ell}$.
\item
\emph{Rule B}: Modify the calculation of coefficients by labelling the clubs with their country and domestic achievement instead of their name. The mathematical formula can remain unchanged but the values are ordered in a decreasing sequence for each association at the end. These coefficients are distributed among the qualified clubs from a given country according to their way of qualification. \\
Formally, suppose that $k$ teams can qualify for \emph{all} European competitions from a given country. The redefined coefficients (associated with the way of qualification instead of the name of the club) determine vector $\mathbf{d} = \left[ d_i \right]$, which is not necessarily non-increasing (e.g.\ the English runner-up may perform better than the English champion even in the long run). The team ranked $\ell$th among the $k$ qualified teams receives the prize associated with the $\ell$th highest element in $\mathbf{d}$.
\end{itemize}
Crucially, both options require the monotonic distribution of coefficients among all qualified teams to ensure incentive compatibility.

As an illustration, see how Rule A works for the English clubs in the 2022/23 season of UEFA club competitions. The first seven teams from this country in the 10-year club coefficient ranking are Chelsea (242), Manchester City (220), Liverpool (215), Manchester United (208), Arsenal (172), Tottenham Hotspur (148), and Leicester City (45). Therefore, in the third pillar of the revenue distribution system, the values $\left[ 242, 220, 215, 208, 172, 148, 45 \right]$ are assigned to the top seven teams in the 2021/22 Premier League. For example, the payment of Arsenal is based on its ``financial coefficient'' of $172$, independently of the identity of the sixth-placed team, whose ``financial coefficient'' is $148$.

Note also that some teams can qualify for European competitions by winning a domestic cup or being the titleholder. They should be considered in the appropriate place of the league ranking to acquire a ``financial coefficient''. As UEFA gives priority to titleholders in seeding rules and to cup winners in filling vacant slots, these teams can essentially occupy the position immediately above/below the first/last club that enters the same/higher-ranked competition. For example, the Italian club Roma has qualified for the 2022/23 UEFA Europa League by winning the 2021/22 UEFA Europa Conference League, and the Europa League berths of Italy are awarded to the fifth-placed team and the cup winner, Roma would receive the fifth-highest UEFA club coefficient of Italian teams. 

Rule B has been recommended by the French mathematician \emph{Julien Guyon} for seeding in the UEFA Champions League \citep{Guyon2015b}, thus, it is only shortly outlined. For instance, four English clubs have entered the 2021/22 Champions League, where they have collected the following number of points: Manchester City (champion, 27), Manchester United (runner-up, 18), Liverpool (third-placed team, 33), and Chelsea (fourth-placed team, 25). Hence, the English champion can be said to have achieved 27 points in this season, and so on. These values are summarised as before, however, they are ordered in a decreasing sequence at the end to guarantee fairness. Otherwise, the runner-up may have a higher coefficient than the champion.

Rule A does not require any fundamental reform and does not affect other systems, while Rule B means a more radical step (but it is able to solve the Champions League seeding problem, too). None of them can account for historic title points, which should depend on the identity of the teams by definition. Thus, the budget available to former titleholders needs to be distributed in a separate pillar that does not influence the share of clubs qualifying due to their latest achievements. For instance, UEFA can allocate 10\% of the current coefficients-based amounts for former titleholders (independently of their performance in the last season since, otherwise, this pillar would not be anonymous), and 90\% for the qualified clubs according to Rule A or B.

To conclude, our note has identified a flaw in the revenue distribution system of the UEFA club competitions: a potential negative ``reward'' for winning a match in a domestic league. Therefore, UEFA is strongly encouraged to choose an incentive compatible allocation mechanism, for example, by adopting one of the proposals above.

\section*{Acknowledgements}
\addcontentsline{toc}{section}{Acknowledgements}
\noindent
We are grateful to \emph{Aleksei Y.~Kondratev} for useful advice. \\
Two anonymous reviewers provided valuable comments on an earlier draft.

\bibliographystyle{apalike}
\bibliography{All_references}

\end{document}